\documentclass[twocolumn]{aastex62}
\usepackage{amsmath}
\graphicspath{{./}{figures/}}

\shorttitle{Impacts of Dust Grains Accelerated by Supernovae on the Moon}
\shortauthors{Siraj \& Loeb}

\begin{document}

\title{Impacts of Dust Grains Accelerated by Supernovae on the Moon}

\email{amir.siraj@cfa.harvard.edu, aloeb@cfa.harvard.edu}

\author{Amir Siraj}
\affil{Department of Astronomy, Harvard University, 60 Garden Street, Cambridge, MA 02138, USA}

\author{Abraham Loeb}
\affiliation{Department of Astronomy, Harvard University, 60 Garden Street, Cambridge, MA 02138, USA}




\begin{abstract}
There is evidence that ejecta from nearby supernovae have rained down on Earth in the past. Supernovae can accelerate pre-existing dust grains in the interstellar medium to speeds of $\sim 0.01 \mathrm{\;c}$. We investigate the survival and impact of dust grains from supernovae on the moon, finding that supernova dust grains can form detectable tracks with widths of $\sim 0.01 - 0.07 \mathrm{\; \mu m}$ and depths of $\sim 0.1 - 0.7 \mathrm{\; mm}$ in lunar rocks. These tracks could potentially shed light on the timings, luminosities, and directions of nearby supernovae.

\end{abstract}

\keywords{Moon -- supernovae: general -- dust -- meteorites, meteors, meteoroids}


\section{Introduction}

There is evidence that nearby supernovae have resulted in the $^{60}\mathrm{Fe}$ and other radionuclides detected in deep-ocean samples \citep{1999PhRvL..83...18K, 2004PhRvL..93q1103K, 2012PASA...29..109F, 2016Natur.532...69W}, the lunar surface \citep{2016PhRvL.116o1104F}, cosmic rays \citep{1974Sci...184.1079R, 2015PhRvL.115r1103K, 2018PhRvD..97f3011K}, and microfossils \citep{2016PNAS..113.9232L}. Supernovae eject can accelerate dust to sub-relativistic speeds $\lesssim 0.1 \mathrm{\; c}$ \citep{1949PhRv...76..583S, 1999APh....12...35B, 2003LNP...598.....W, 2015ApJ...806..255H}. Studying dust accelerated by supernovae could elucidate the history of nearby supernovae \citep{2004PhRvL..93q1103K, 2016ApJ...826L...3T, 2016Natur.532...69W, 2017ApJ...840..105M} and constrain theoretical models of supernovae \citep{2015MNRAS.446.2089W, 2017MNRAS.465.4044B, 2017MNRAS.465.3309D, 2019MNRAS.489.4465K}. Searches on the lunar surface for signatures from interstellar objects have been suggested \citep{2019arXiv190808543S, 2019arXiv190705427L}. Dust grains accelerated by supernovae could appear as meteors in the Earth's atmosphere \citep{2020arXiv200201476S} and as chemical enrichments in subsurface layers on the Moon \citep{2017hsn..book.2507C}. Given that NASA's Artemis program plans to  establish a sustainable base on the Moon  by 2024\footnote{https://www.nasa.gov/specials/artemis/}, it is now particularly timely to explore the detection signatures of interstellar dust on its regolith and rocks. In this \textit{Letter}, we explore the flux of dust grains accelerated by supernovae at the lunar surface and the expected rate of the resulting impact tracks in lunar materials.

Our discussion is structured as follows. In Section \ref{sec:survival}, we consider the survival of dust grains accelerated by supernovae as they travel to the Moon. In Section \ref{sec:rate}, we investigate the effect of radiation pressure from supernovae on the acceleration of dust grains. In Section \ref{sec:rate}, we explore the lunar impact rate of such dust grains. In Section \ref{sec:depth}, we estimate the depth to which these dust grains penetrate lunar materials. In Section \ref{sec:area}, we compute the expected track densities in lunar rocks. Finally, in Section \ref{sec:disc} we summarize key predictions and implications of our model.


\section{Grain Survival}
\label{sec:survival}

Coulumb explosions from charge accumulation hinder the distance that supernova dust grains can travel in the interstellar medium (ISM). The surface potential $\phi_{max}$ above which the grain will be disrupted by Coulomb explosions, assuming a typical tensile strength of $\sim 10^{10} \; \mathrm{dyn \; cm^{-2}}$, is \citep{2015ApJ...806..255H},

\begin{equation}
    e \phi_{\mathrm{max}} \simeq \mathrm{0.1 \; keV} \left(\frac{r}{\mathrm{0.01 \; \mu m}} \right) \; \; ,
\end{equation}
where $e$ is the electron charge. There is also a maximum surface potential $U_{\mathrm{max, H}}$ due to the fact that a large potential can halt electrons from overcoming the surface potential and therefore escaping the grain \citep{2017ApJ...848...31H},

\begin{equation}
    e U_{\mathrm{max,H}} = 2 m_e v^2 \simeq 0.1 \mathrm{\; keV} \left( \frac{v}{0.01 \mathrm{\; c}} \right)^2 \; \; ,
\end{equation}
where $m_e$ is the electron mass. Combining equations (1) and (2),

\begin{equation}
    \label{eq:33}
    \left(\frac{r}{0.01 \mathrm{\; \mu m}} \right)\gtrsim  \left( \frac{v}{0.01 \mathrm{\; c}} \right)^2 \; \; ,
\end{equation}
yields a minimum grain radius of $r \sim 0.01 \mathrm{\; \mu m}$ for survival at a speed of $v \sim 0.01 \mathrm{\; c}$ and implies that grains above this size will not undergo Coulomb explosions while traveling through the ISM.

Significant slow-down for dust grains traveling through the ISM occurs at distance where the total momentum transferred by particles in the ISM to the object is comparable to the initial momentum of the object. However, at the speed of $\sim 0.01 \mathrm{\; c}$, the stopping distance of a proton in silicate material is comparable to the size of the grain for a radius $r \sim 0.4 \mathrm{\; \mu m}$ (Figure 2 of \citealt{2017ApJ...837....5H}), so the stopping distance through the ISM for $r \lesssim 0.4 \mathrm{\; \mu m}$ and $v \sim 0.01 \mathrm{\; c}$ is,
\begin{equation}
    d_{ISM} \sim 250 \mathrm{\; pc} \left( \frac{\rho}{3 \mathrm{\; g \; cm^{-3}}} \right) \left( \frac{n_p}{0.1 \mathrm{\; cm^{-3}}} \right)^{-1} \; \; ,
\end{equation}
given local proton density of the ISM is $n_p \sim 0.1 \mathrm{\; cm}^{-3}$ \citep{2011ARA&A..49..237F}. At faster speeds, $d_{ISM} \gtrsim 250 \mathrm{\; pc}$, but this contribution does not significantly enhance the flux of dust grains from supernovae, so it is not studied here.

Additionally, thermal sublimation in both the ISM and the solar radiation field and Coulomb explosions in the solar wind do not limit further the sizes and speeds of grains considered here \citep{2015ApJ...806..255H}. We note that interstellar dust particles of similar sizes considered here that travel at typical speeds are excluded from the inner solar system due to heliospheric and radiation pressure effects (see \citealt{2019SSRv..215...43S} for a review on interstellar dust in the solar system).

\section{Grain Acceleration}
\label{sec:acceleration}

Radiation pressure from a supernova accelerates dust grains to sub-relativistic speeds \citep{1949PhRv...76..583S, 1999APh....12...35B, 2003LNP...598.....W, 2015ApJ...806..255H}.  Assuming a typical bolometric luminosity of $\sim 10^{8} \; L_{\odot}$, equation (12) in \citet{2015ApJ...806..255H} yields,

\begin{equation}
    v \sim 0.01 \mathrm{\;c} \left(\frac{d}{\mathrm{5 \times 10^{18} \; cm}} \right)^{-1/2} \left(\frac{r}{\mathrm{0.01 \; \mu m}} \right)^{-1/2} \; \; ,
    \label{eq:111}
\end{equation}
where $r$ is the grain radius, $d$ is the initial distance, and $v$ is the grain speed. Since dust grains are sublimated within a distance of $d \sim 10^{16} \; \mathrm{cm}$, equations \eqref{eq:33} and \eqref{eq:111} jointly constrain the range of sizes that this method is applicable to grains of size $\sim 0.01 - 0.07 \mathrm{\; \mu m}$, corresponding to speeds of $\sim 0.01 - 0.03 \mathrm{\; c}$.

Adopting the canonical dust mass fraction in the ISM of $\sim 0.01$, $\sim 2 \times 10^{-4} \; M_{\odot}$ of pre-existing dust around each supernova exists within a distance of $\; 5 \times 10^{18} \mathrm{cm}$, interior to which $r \sim 0.01 \mathrm{\; \mu m}$ dust grains can be accelerated to $\sim 0.01 \mathrm{c}$. The fraction of ISM dust with size $r \gtrsim 0.01 \mathrm{\; \mu m}$ is $\sim 0.5$ \citep{1977ApJ...217..425M}, so $\sim 10^{-4} \; M_{\odot}$ of dust is accelerated per supernova, corresponding to $\sim 10^{44}$ grains. Additionally, the size distribution of ISM dust grains follows a power law with exponent -3.5 \citep{1977ApJ...217..425M}. 

Dust grains produced in supernovae travel at the ejecta speed of $\sim 0.01 \mathrm{\; c}$ because the drag timescale due to gas \citep{2015A&A...575A..95S} is short relative to the acceleration timescale by radiation pressure \citep{2017ApJ...837....5H}. However, the abundance and size distribution of supernova ejecta dust are model-dependent, unlike pre-existing dust grains in the ISM. The ejecta model for typical Type II-P supernovae, presented in Figure 4 of \cite{2015A&A...575A..95S}, peaks for grain radii of $r \sim 0.02 \; \mathrm{\mu m}$, a size bin at which $\sim 5 \times 10^{-3} \mathrm{\; M_{\odot}}$ of dust grains is produced per supernova, which also corresponds to $\sim 10^{44}$ grains.  Further modelling of the ejecta of core-collapse supernovae in general will reveal the true abundance of sub-relativistic dust grains produced relative to pre-existing ISM dust that is accelerated by the radiation pressure of the supernova. Additionally, ejecta-formed dust likely undergo different dynamics relative to accelerated ISM dust grains \citep{2018arXiv180106859F, 2019BAAS...51c.410F}.

In addition, massive stars such as luminous blue variables can reach stellar wind mass-loss rates of $\sim 10^{-5} \; M_{\odot}\mathrm{ \; yr^{-1}}$ \citep{2011AJ....142...45A}, which could lead to $\sim 10^{-4} \; M_{\odot}$ of pre-existing dust that would be accelerated to $\gtrsim 0.01 \mathrm{\; c}$ due to radiation pressure from the supernova. However, luminous blue variables are a small fraction of all supernovae, so we do not consider them here.

\section{Impact rate}
\label{sec:rate}

The local timescale between core-collapse supernovae is estimated to be $\tau_{SN} \sim 2 \mathrm{\; Myr}$ within a distance $d_{SN} \sim 100 \mathrm{\; pc}$ \citep{2004PhRvL..93q1103K, 2016ApJ...826L...3T, 2016Natur.532...69W, 2017ApJ...840..105M}, implying a timescale of $\tau_{SN} \sim 0.1 \mathrm{\; Myr}$ within a distance $d_{SN} \sim 250 \mathrm{\; pc}$.

Type II supernovae are an order of magnitude more common than Type Ib/c supernovae, and so we focus our discussion on them \citep{2007ApJ...657L..73G}.

The impact area density on the lunar surface due to one supernova at $\sim 250 \mathrm{\; pc}$ of $r \sim 0.01 \; \mathrm{\mu m}$ dust grains traveling at $\sim 0.01 \mathrm{\; c}$ accelerated by core-collapse supernovae is therefore,

\begin{equation}
    \begin{aligned}
    \sigma_{SN} \sim 10^3 \; \mathrm{cm^{-2}} \left( \frac{d_{SN}}{250 \; \mathrm{pc}} \right)^{-2} \left( \frac{\tau_{SN}}{0.1 \; \mathrm{Myr}} \right) \; \; .
    \end{aligned}
\end{equation}

\section{Impact Tracks}
\label{sec:depth}
Lunar surface material is composed primarily of silicates \citep{1997GeCoA..61.2331K, 2006JGRE..11112007P, 2007M&PS...42.2079M, 2010Sci...329.1507G}, and so we adopt the properties of quartz for our impact depth penetration analysis. Since the impacts occur at sub-relativistic speeds, we consider the dust grains as collections of constituent nuclei (Si, as a fiducial example). The stopping power $dE/dx$ as a function of Si nucleus speed for impacts in quartz is adopted from Figure 2 of \citet{2017ApJ...837....5H}, allowing the penetration depth $l$ of a single ion to be computed.

The sideways shock moves at a speed $\lesssim 10^{-2}$ of the dust speed, producing a `track' is in which the depth greatly exceeds the width, which is comparable to the size of the dust grain. Only a fraction of the total energy is shared with the target per penetration depth of a single ion. This energy fraction $\epsilon$ is found by lowering the sideways speed $v_{shock}$ per penetration depth $x$ such that the timescale to traverse the dust grain radius $r$ sideways, $r/v_{shock}$, equals the timescale to traverse the penetration depth $l$ at the dust speed, $l/v$. This means that $v_{shock} = (rv/l)$, implying an energy deposition fraction, $\epsilon = r/l$. Since we deposite the fraction $\epsilon$ of the dust energy per penetration depth $x$, the total penetration depth is, $D \sim l^2/r$.

\begin{figure}[h]
  \centering
  \includegraphics[width=0.9\linewidth]{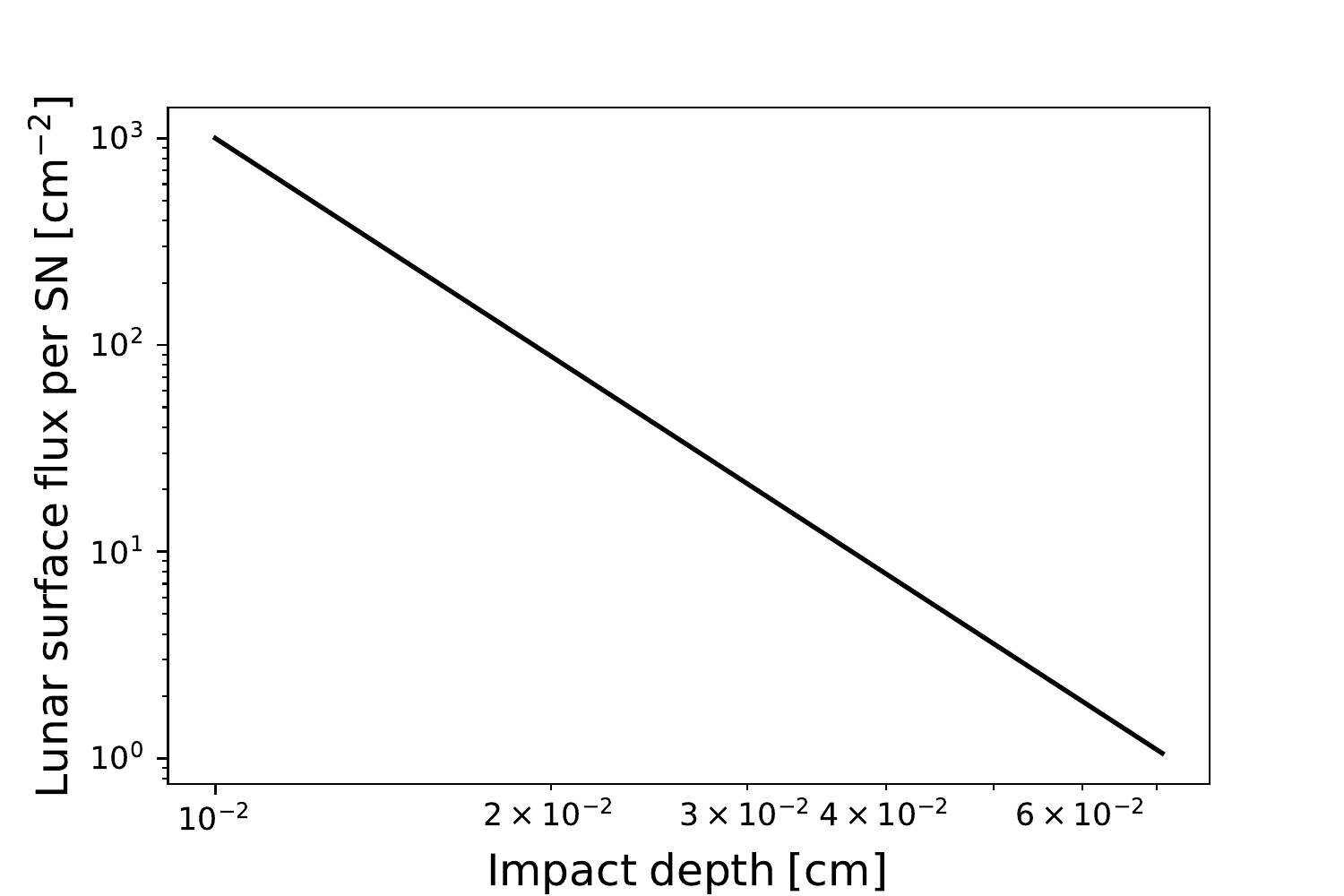}
    \caption{Lunar surface flux of dust grains as a function of impact depth, accelerated by a single supernova at $\sim 250 \mathrm{\; pc}$. The range corresponds to grain sizes of $\sim 0.01 - 0.07 \; \mathrm{\mu m}$ and speeds of $\sim 0.01 - 0.03 \; \mathrm{c}$.
}
    \label{fig:dependence}
\end{figure}

With a grain radius of $r \sim 0.01 \mathrm{\; 
\mu m}$ and a penetration depth per ion of $l \sim 1 \mathrm{\; \mu m}$ the total penetration depth is $D \sim 0.1 \mathrm{\; mm}$. The penetration depth for a constant size scales as $D \propto v^4$, since $dE/dx$ is constant to order unity over the speeds considered; however, due to the size-speed constraint of equation \eqref{eq:33} and the fact that $D \propto r^{-1}$, the penetration depth actually scales as $dD/dv \propto v^{2}$. The abundance of dust grains, $N$, scales as the volume ($d^3$) multiplied by the ISM dust grain size distribution ($r^{-3.5}$). Since equation \eqref{eq:111} yields $rd \propto v^{-2}$ and equation \eqref{eq:33} gives $r \propto v^2$, in total the abundance scales as $dN/dv \propto v^{-5}$. As a function of depth, the abundance scales as $dN/dD \propto v^{-7}$, which, using equation \eqref{eq:111}, yields a total dependence of the dust grain flux at the lunar surface of $dN/dD \propto D^{-7/2}$, as indicated in Figure \ref{fig:dependence}.

\section{Signatures in lunar materials}
\label{sec:area}
The lunar regolith is steadily overturned by micrometeroid impacts (see \citealt{2011P&SS...59.1672G} for a review of the lunar dust environment). The approximate depth $D$ to which the lunar regolith is overturned as a function of timescale $\tau$, given in Figure 9 of \citet{2018Icar..314..327C}, is,

\begin{equation}
D \sim 10 \mathrm{\; cm} \left( \frac{\tau}{0.1 \mathrm{\; Myr}} \right)^{1/4} \; \; ,
\end{equation}
which implies that tracks resulting from the penetration of dust grains accelerated by nearby supernovae into the lunar regolith cannot be discovered, given that the overturned depth between supernovae within $\sim 250 \mathrm{\; pc}$ is a thousand times larger than the track depth for $r \sim 0.01 \mathrm{\; 
\mu m}$ dust grains at $\sim 0.01 \mathrm{\; c}$.

\begin{figure}[h]
  \centering
  \includegraphics[width=0.9\linewidth]{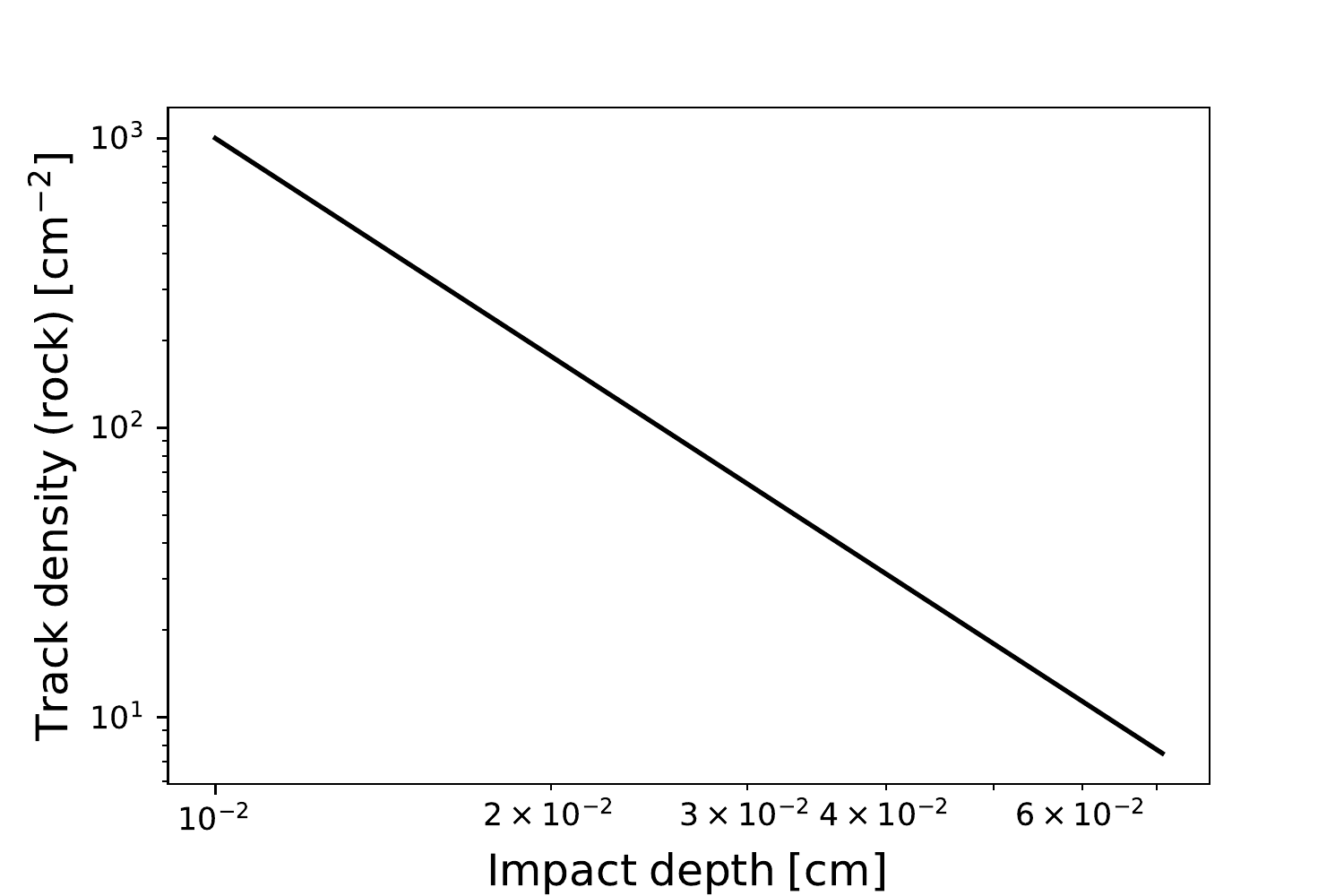}
    \caption{Current expected track density as a function of track depth in the lunar rock, resulting from impacts of dust grains with sizes of $\sim 0.01 - 0.07 \; \mathrm{\mu m}$ and speeds of $\sim 0.01 - 0.03 \; \mathrm{c}$. In contrast with Figure \ref{fig:dependence}, this plot is corrected for the lunar rock erosion rate discussed in the text.
}
    \label{fig:regolith}
\end{figure}

Lunar rocks, on the other hand, are eroded by micrometeroids at a rate of $\sim 0.1 - 1 \mathrm{\; mm \;  Myr^{-1}}$ \citep{1972IAUS...47..330C, 1973LPI.....4..558N, 1977NASSP.370..585F, 1995LPI....26.1055N, EUGSTER20033, 2006mess.book..829E}. We conservatively adopt a rate of $\sim 1 \mathrm{\; mm \;  Myr^{-1}}$. They have surface lifetimes of $1 - 50 \mathrm{\; Myr}$ \citep{1980asfr.symp...11W, 1991lsug.book.....H}. Impact tracks can be discovered in lunar rocks at rates shown in Figure \ref{fig:regolith}.

Micrometeoroid impacts result in craters with depths comparable to their widths, and cosmic rays develop tracks with widths of $\sim 1 \mathrm{\; pm}$, so sub-relativistic dust grain tracks are uniquely identifiable by their characteristic $\sim 0.01 \; \mu m$ widths and depths that greatly exceed what would be expected from a micrometeoroid impact.

\section{Discussion}
\label{sec:disc}

We demonstrated that tracks resulting from $r \sim 0.01 - 0.07 \mathrm{\; \mu m}$ dust grains accelerated by supernovae to speeds of $r \sim 0.01 - 0.03 \mathrm{\; c}$ can be discovered in the lunar rocks. Studies of lunar rocks could shed light on the history of supernovae within the past $\sim 1 \mathrm{\; Myr}$ and within $\sim 250 \mathrm{\; pc}$, with the potential to reveal the timings, lumonsities, and directions of recent supernovae. The expected density of tracks is $\sim 10^{3} \; \mathrm{cm^{-2}}$ for depths of $\sim 0.1 \mathrm{\; mm}$ and widths of $\sim 0.01 \mathrm{\; \mu m}$ and $\sim 1 \; \mathrm{cm^{-2}}$ for depths of $\sim 0.7 \mathrm{\; mm}$ and widths of $\sim 0.07 \mathrm{\; \mu m}$. 
However, since lunar rocks may be covered by regolith material with typical grain sizes of $\sim 100 \mathrm{\; \mu m}$ \citep{1991lsug.book.....H}, impacts of the sub-relativistic grains considered here could catastrophically disrupt such grains instead of forming tracks within a rock, potentially reducing the actual density of tracks observed on lunar rocks.

The six Apollo missions brought to Earth 2200 lunar rocks\footnote{https://curator.jsc.nasa.gov/lunar/} which could be searched these tracks. While cosmic ray tracks have been discovered in lunar samples \citep{1974GeCoA..38.1625D, 1980asfr.symp..331C, 1981InEPS..90..359B}, these tracks would be differentiable by their widths which would be at least an order of magnitude larger. Extraterrestrial artifacts, such as microscopic probes akin to Breakthrough Starshot\footnote{https://breakthroughinitiatives.org/initiative/3}, could also form such tracks.

\section*{Acknowledgements}
We thank Brian J. Fry and Thiem Hoang for helpful comments. This work was supported in part by a grant from the Breakthrough Prize Foundation. 





\bibliography{bib}{}
\bibliographystyle{aasjournal}



\end{document}